%% file: aaai.tex
\def\year{2020}\relax
\documentclass[letterpaper]{article} 
\usepackage{aaai20}  
\usepackage{times}  
\usepackage{helvet} 
\usepackage{courier}  
\usepackage[hyphens]{url}  
\usepackage{graphicx} 
\usepackage{graphicx}  

\usepackage{booktabs} 
\usepackage{xspace}
\usepackage{url}
\usepackage{listings}
\usepackage{bigstrut}
\usepackage{subcaption}
\usepackage{graphicx}
\usepackage{color}
\usepackage{amssymb}
\usepackage{amsmath}
\usepackage[ruled,vlined,linesnumbered]{algorithm2e}
\usepackage{soul}
\usepackage[htt]{hyphenat}
\usepackage{caption}

\definecolor{lightgray}{rgb}{.9,.9,.9}
\definecolor{darkgray}{rgb}{.4,.4,.4}
\definecolor{purple}{rgb}{0.65, 0.12, 0.82}
\definecolor{orange}{rgb}{1,0.5,0}

\newcommand{\point}[1]{\vspace{.05in} \par\noindent\textbf{#1}. }

\newcommand{\squishlist}{
   \begin{list}{$\bullet$}
    { \setlength{\itemsep}{0pt}      \setlength{\parsep}{3pt}
      \setlength{\topsep}{3pt}       \setlength{\partopsep}{0pt}
      \setlength{\leftmargin}{1.0em} \setlength{\labelwidth}{1em}
      \setlength{\labelsep}{0.5em} } }

\newcommand{\squishlisttwo}{
   \begin{list}{$\bullet$}
    { \setlength{\itemsep}{0pt}    \setlength{\parsep}{0pt}
      \setlength{\topsep}{0pt}     \setlength{\partopsep}{0pt}
      \setlength{\leftmargin}{2em} \setlength{\labelwidth}{1.5em}
      \setlength{\labelsep}{0.5em} } }

\newcommand{\squishend}{
    \end{list}  }

\newcommand{\tool}{\textsc{$\text{A}\textsuperscript{4}$}\xspace}

\definecolor{OliveGreen}{rgb}{0,0.6,0}

\lstdefinelanguage{JavaScript}{
  keywords={typeof, new, true, false, catch, function, return, null, catch, switch, var, if, in, while, do, else, case, break},
  keywordstyle=\color{blue}\bfseries,
  ndkeywords={class, export, boolean, throw, implements, import, this},
  ndkeywordstyle=\color{darkgray}\bfseries,
  identifierstyle=\color{black},
  sensitive=false,
  comment=[l]{//},
  morecomment=[s]{/*}{*/},
  commentstyle=\color{purple}\ttfamily,
  stringstyle=\color{red}\ttfamily,
  morestring=[b]',
  morestring=[b]"
}

\title{\tool: Evading Learning-based Adblockers}
\begin{document}

\maketitle

\begin{abstract}
Efforts by online ad publishers to circumvent traditional ad blockers towards regaining fiduciary benefits, have been demonstrably successful. As a result, there have recently emerged a set of adblockers that apply machine learning instead of manually curated rules and have been shown to be more robust in blocking ads on 
websites including many social media sites such as Facebook \cite{adblockai}. 
Among these, AdGraph \cite{iqbal2020adgraph} which uses rich contextual information to perform classifications, is arguably, the state of the art learning-based adblocker. In this paper, we develop \tool, a tool that intelligently crafts adversarial samples of ads to evade AdGraph. Unlike the popular research on adversarial samples against images or videos that are considered less- to un-restricted, the samples that \tool generates preserve application semantics of the web page, or are \textit{actionable}. Through several experiments we show that \tool can bypass AdGraph about \underline{60\%} of the time, which surpasses the state-of-the-art attack by a significant margin of \underline{84.3\%}; in addition, changes to the visual layout of the web page due to these perturbations are imperceptible. We envision the algorithmic framework proposed in \tool is also promising in improving adversarial attacks against other learning-based web applications with similar requirements.
\end{abstract}

\input{docs/intro}
\input{docs/background}
\input{docs/related_work}
\input{docs/system}
\input{docs/evaluation}
\input{docs/discussion}
\input{docs/conclusions}

\bibliographystyle{aaai}
\bibliography{ref}

\end{document}

%% file: docs/intro.tex
\section{Introduction}
\label{sec-1}
%
%

As adblockers have gained popularity in recent years \cite{pagefairreport2017}, 
ad publishers have started fighting back towards recovering their revenues.
Specifically, many techniques have emerged towards circumventing
the current generation of adblockers \cite{zhu2018measuring}.
Notably, prior work \cite{zhu2018measuring} has shown that from among Alexa's top 10K websites, including many social media sites, more than 30\% have client-side JavaScript code that serve as countermeasures against adblocker use.
%
%

Conventionally, adblockers rely on manually curated (and maintained) blacklists,
with rules/signatures that are matched against resource request URLs sent from the
browser and the elements rendered in a web page.
Unfortunately, maintaining such blacklists does not scale and is error-prone.
Furthermore, they are also fairly easy to subvert (just as antivirus signatures)~\cite{mughees2017detecting,iqbal2017ad}.
Given these limitations, there has been a recent trend towards the emergence of machine-learning-based adblockers \cite{iqbal2020adgraph,storey2017future,tigas2019percival} towards improving the effectiveness and accuracy over signature-based adblockers.
Such adblockers can be categorized into ``perceptual'' and ``non-perceptual'' classes.
Perceptual adblockers \cite{storey2017future,tigas2019percival} block ads by recognizing
visual cues (e.g. "sponsored" or other marketing keywords) in the web page.
It is claimed that these are more robust because some regulators (e.g. FTC) require publishers to disclose
the ``ad nature'' of their online content.
However, recent research has shown that these vision-based adblockers can be easily fooled by adversarial examples; this is an artifact of the recent advances in adversarial machine learning (AML)~\cite{tramer2018ad}.
In brief, by only adding human-imperceptible perturbation pixels to the ad images, a classifier can be fooled.

In contrast, non-perceptual adblockers detect ads based on non-visual features such as the URL contents and page structure.
The state-of-the-art of non-perceptual ML-based adblocker, arguably, is AdGraph \cite{iqbal2020adgraph} \footnote{\cite{sjosten2019filter} is a concurrent effort on extending AdGraph to automatically generate filter lists but its classification pipeline code has not been released. 
Because of this our work targets only AdGraph; however, we anticipate that one can draw similar conclusions given the similarity in its design to AdGraph.}.
Unlike most previous works that rely on URL text or code structure alone, AdGraph builds
a page loading graph for a web page with rich contextual information, and extracts features from it to classify HTTP(S) request nodes.
The claim is that AdGraph's use of fine-grained structural/contextual information
provides improved robustness over other adblockers,
since it requires significant changes to a web page to convincingly alter the structure towards
concealing the ad.
Furthermore, the use of non-visual features
inhibit the applicability of adversarial attack techniques from the unconstrained domain (i.e., image) \cite{tramer2018ad}. 
In a nutshell, a major contribution in this paper is that we show neither of these conclusions holds true with AML.
%
%

While there has been success on applying adversarial examples
in unconstrained domains (e.g., images) \cite{carlini2017towards},
the feasibility of crafting such inputs in domains with more stringent constraints (e.g., web pages)
remains largely unexplored. In the constrained web domain, the visual perceptibility is altered if
the semantics of the web page are changed.
%
%
Specifically,
web pages are processed by browsers prior to user exposition (unlike images). Thus, rather than
the magnitude of the perturbation being the most important criterion, what matters is that the rendered web page after applying the perturbation presents the same look-and-feel and functionality to the user.
This requirement forces the perturbations to be what we call {\em actionable} and
requires a rethinking of what constraints must be enforced while crafting adversarial
samples.
Extending this principle to (ML based) adblockers, given the goal of perturbing an ad resource request to bypass the ML detection model: (i) the adversarial example should be actionable in that it must be ``mapped back'' to the appropriate valid web page
and (ii) the modified request must preserve its original functionality of directing the requester to the remote ad server
i.e., this requires the ``functional'' parts of the page to be equivalent before and after modification; only the other ``non-functional'' parts can be perturbed to fool the classifier.
Our goal in this paper is to develop a tool (we call this \tool) to automatically craft such
{\em actionable} perturbations that can subvert AdGraph.
%
%
Towards achieving this, our challenge is to realize the
two properties below.

\textbf{Feature-space actionability}: First, any perturbation that is generated in the feature-space must be bounded by domain-specific constraints.
Such constraints can be expressed explicitly by a set of mathematical formulas that the perturbed feature vector must comply with, so as to preserve the functionality of the original web resource and the validity of the page. Since these constraints are typically manually identified (depending on the functionality
to be preserved), the attacker needs to find a way to integrate them properly into the adversarial example generation algorithm.

\textbf{Application-space actionability}: Second, upon mapping the feature-space perturbations back to
the raw input space (web page), the computed modifications may cause undesired changes
either because they violate some structural constraints (e.g., the number of nodes in a DOM tree
cannot be negative) or the violation of
complicated inter-relationships between objects in the page, that are projected to the extracted feature values.
Correcting these changes will require adding offsets to the adversarial perturbations 
from the feature-space. However,
blindly applying them may render
them non-adversarial, i.e., the attacker must
account for these offsets when generating perturbations.

As our primary contribution, we design \tool - \textbf{A}ctionable \textbf{A}d \textbf{A}dversarial \textbf{A}ttack to generate perturbations that are actionable both in the feature-space and 
the application-space. \tool only requires minimal domain knowledge towards providing a set of \textit{seed} features that can be mapped from feature space back to the input or application space.
Specifically, it has the following desirable characteristics.

\squishlist
    \item \textbf{Efficient crafting:} Inspired by the popular gradient-based attack, Projected Gradient Descent (PGD) \cite{kurakin2016adversarial}, \tool \textit{iteratively} 
accounts for the unique constraints of the web environment, to generate potent adversarial samples. 
Our evaluations show that \tool achieves a success rate of about 60\% i.e.,  it subverts 60\% 
of the inputs that were originally classified as ad/tracker by AdGraph, to benign. In comparison, a baseline attack that only accounts for actionability requirements for one iteration (non-iteratively) can only achieve a success rate of less than 33\%, while a weaker baseline cannot generate any viable example.
    \item \textbf{Actionability:} All perturbed web resources are guaranteed to comply with both
the feature-space and application-space constraints. Such compliance makes these examples practical in
that they still carry out their ad/tracker functionalities.
    \item \textbf{Stealthiness:} \tool generates perturbations that have low detectability. In our setup, generated perturbations are bounded and concealed with respect to the corresponding web pages, which make them hard to be detected by adblockers; furthermore, they
are invisible/imperceptible to users (except for displaying the ads).
\squishend



%% file: docs/background.tex
\section{Background and Related work}
\label{sec-2}
In this section, we provide brief background on adblockers and AML.
We also discuss relevant related work.

\point{Non-perceptual ML-based Adblocking}
Because rule-based adblockers are plagued by scale/errors and demonstrable attacks, 
machine learning (ML)-based adblockers are emerging.
Instead of relying on hard-coded blacklists that are manually curated and maintained, they
use ML to infer the intrinsic patterns of ad/tracker resources from different representations of web pages.
Previously, URL strings and JavaScript code have been used 
as features to represent web resources in ML models \cite{bhagavatula2014leveraging}.
However, these attempts have low accuracy because the representations used are incomplete in capturing 
the distinguishing characteristics of ad and non-ad resources.
This led to AdGraph \cite{iqbal2020adgraph}, a more recent work on identifying ad resources using ML; 
(arguably) because this is the state-of-the-art in this field, we 
consider this as our target in this paper.

\point{AdGraph}
By instrumenting the browser core, AdGraph collects a comprehensive set of browser-internal events to stitch together a graph that represents the interactions among the HTML page elements, network requests, and JavaScript executions (e.g., a web element is dynamically created by a script).
This representation is then used to train a classifier for identifying advertising and tracking resources.
With support from this rich loading context, AdGraph extracts 65 features from a resource load, and classifies the request based on these features.
These features can be categorized into two types: structural and content-based.
Content-based features include (but not limited to) certain susceptible ad-related keywords in the URL and the requested resource type (e.g. image, iframe).
AdGraph's classifier uses Random Forest as the underlying model, which is non-differentiable.
As discussed later in \S \ref{sec-4}, this choice hinders traditional AML based attacks as they require gradient information to guide the adversarial example generation.
Moreover, from the 65 features AdGraph uses, 5 of them are categorical i.e., will be converted into more than 250 sparse one-hot-encoded features.
Such sparsity not only poses new challenges for existing adversarial attacks that expect dense data, 
but require additional constraints to ensure the validity of the one-hot vectors
(we discuss how \tool overcomes these in \S \ref{sec-4}).

%% file: docs/related_work.tex
\point{Adversarial attacks on ML models}
The common setup for adversarial attacks against binary ML classifiers is that given a model and an input that is classified as malicious, an attacker needs to modify the input to flip the model's classification result.
Formally, suppose a classifier defined by its prediction function $P_{model}$ and an input $x$ with its malicious label $l_{mal}$; an attacker needs to find an adversarial transformation $T_{adv}$ such that $P_{model}(T_{adv}(x_{input})) \neq l_{mal}$.
The AML community defines different levels of model transparency to describe the knowledge that an attacker possesses with regards to the target classifier:
\squishlist
    \item With \textbf{White-box attacks}, an attacker is assumed to know all the information about the model, including but not limited to the model internals (e.g., the classifier model type, parameters), the training dataset and feature definitions.
    \item With \textbf{Grey-box attacks}, the attackers do not know the internals of the model, but know the training dataset and feature definitions. Further, the attacker can query the target classifier about the label for a specific input.
\squishend

\point{Gradient-based attacks}
One popular attack is based on the Fast Gradient Sign Method (FGSM) \cite{bruna2013intriguing}, which leverages the gradients derived from the target classifier to compute the perturbation that maximizes its loss function with respect to the particular malicious input. 
Given the loss function of the target model $L_{model}$, FGSM computes its perturbation $\eta$ as $\eta = \epsilon * sign(\nabla_x L_{model})$, where $\epsilon$ is the norm constraint specified by the attacker.
%
There are also other variants \cite{dong2018boosting} that follow the "loss-maximizing" philosophy used in FGSM. They are generally referred to as gradient-based attacks.
Since these attacks all use the gradient information from the target model, they should be considered as white-box attacks. 

Gradient-based attacks generate perturbations that are bounded based on different $L_p$ ($\epsilon$ above) norms (e.g. $L_0$, $L_2$ or $L_{inf}$).
%
These traditional norms, bounds, or thresholds (referred to as norms in the paper) measure the magnitude of the perturbation, and are thus primarily suitable for visual domain applications (lower norms 
generally mean less visually-detectable changes) wherein  human imperceptibility is the auxiliary
characteristic desired in a perturbation.
%
%
%
In the web space however, the perturbed page has complex structures and is processed by the browser which parses and renders the page. 
Thus, the norms can no longer capture what is a ``desirable perturbation,'' and do not work well. 
%
In other words, 
new metrics are needed to effectively capture the properties of functionality preservation and stealthiness of the perturbed web page.

\point{Projected Gradient Descent}
Being a single-step attack, FGSM suffers from low success rates, especially when gradients 
cannot provide sufficiently accurate guidance (usually the case for non-white-box attacks). 
%
%
One can improve the success rate by applying FGSM iteratively; this is known as 
the Basic Iterative Method, or Projected Gradient Descent \cite{kurakin2016adversarial}.
Essentially, PGD performs FGSM multiple times with a smaller step-size, or $\alpha$.  
Formally, the search procedure can be expressed as:
\begin{equation} \label{equation:pgd}
\begin{gathered}
    x_0 = x_{input} \\
    x_{n+1} = Clip_{\epsilon}(x_{n} + \alpha * sign(\nabla_x L_{model})) 
\end{gathered}
\end{equation}
where, for a given a given input vector $A$, 
\begin{equation} \label{equation:clipping}
Clip(A_i, \epsilon)=
\begin{cases}
A_i-\epsilon, & \text{if } A_i < A_i-\epsilon\\
A_i+\epsilon, & \text{if } A_i > A_i+\epsilon\\
A_i, & {otherwise.}
\end{cases}
\end{equation}
%
%
\tool is inspired by the iterative philosophy used in PGD, and extends its simple clipping mechanism to an extensive feedback loop (\S \ref{sec-4}), which seeks to produce actionable perturbations targeting ML-based adblockers.

%% file: docs/system.tex
\section{\tool: Actionable Ad Adversarial Attack}
\label{sec-4}
%
In this section, we describe how \tool crafts actionable (both in the feature and application spaces
as discussed in \S~\ref{sec-1}) and stealty adversarial examples in web domain.
%
%
Recall that being actionable in the above two spaces refer to the following (i) in the feature space, explicit numeric constraints defined based on domain knowledge, in order to maintain the validity, functionality and stealthiness of the ad request, must be complied with by its perturbed adversarial feature vector; and (ii) in the application space, the perturbed feature vector must be successfully mappable back to the original web page.
%
We point out that actionable perturbations in the feature-space are not \textit{naturally} actionable in the application-space; implicit/unpredictable side-effects that could result when the feature-space perturbations are mapped back to application-space (e.g. a feature space perturbation of adding nodes to a page changes other features such as the average connection degree) must be considered when generating perturbations,
by \tool.
%
%

%
\subsection{Threat Model}
\label{section:threat-model}
\begin{figure}[htbp]
\centering
\includegraphics[width=0.65\columnwidth]{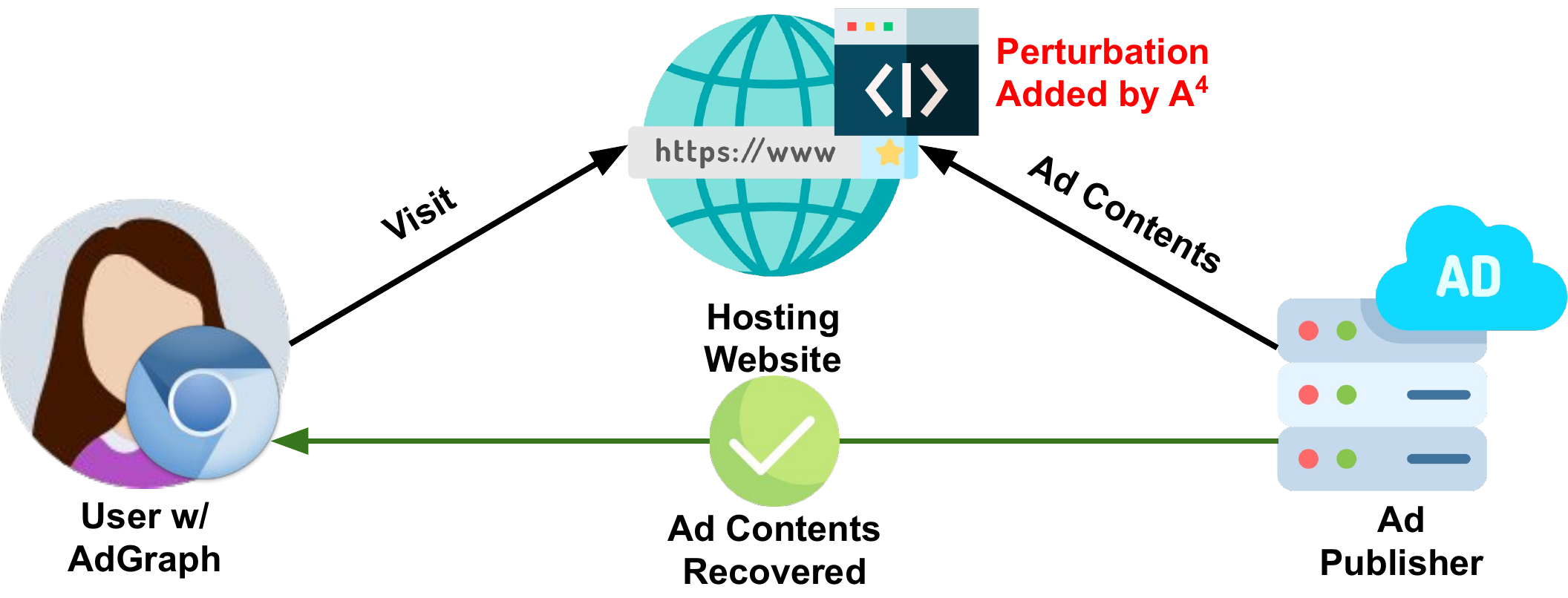}
\caption{Different participants in \tool's threat model}
\label{figure:threat-model}
\end{figure}

Before diving into the details of \tool's algorithm, we first define our threat model.
As mentioned in \S \ref{sec-2}, AdGraph is a full-fledged web browser with custom modifications for blocking ad/tracker resources.
Generally, there are three participants when a user visits a website using AdGraph: a user, a hosting website and an ad publisher;
their relationships are depicted in Figure \ref{figure:threat-model}.
As shown, the objective of \tool is to help the website recover its ad revenue lost due to ads getting blocked by AdGraph.
\tool achieves this by adding perturbations to content generated by the hosting website, so that the classifier used by AdGraph is fooled into mis-classifying the ad resource as a non-ad.
We assume a grey-box attack setup as elaborated in \S \ref{sec-2}, because on one hand, even though AdGraph has been open-sourced, it is easy for other ML-based adblockers to hide their model internals; on the other hand, the web is a public space and datasets collected from it can be conveniently crawled and replicated in practice.

Recall that \tool is a gradient-based attack which requires the knowledge of model internals.
However, AdGraph uses Random Forest (RF) as its classifier which is non-differentiable (i.e. no gradient can be computed),
and thus, we need to find a way to estimate the gradients.
Moreover, although AdGraph itself has been open sourced publicly \cite{iqbal2020adgraph}, we do not want to limit \tool to complete white-box setups only,
which means that even if the target model is differentiable, its gradients can be inaccessible (e.g., when only its prediction APIs are exposed).
Thus, we assume a grey-box threat model and make \tool a transfer-based attack 
%
where the attacker has no access to any model internals (model type, gradients etc.), but is aware of the training dataset and feature definitions
(see \S \ref{sec-2}).
Based on the above, our perturbations should meet the following requirements from the practicality/usability perspective:
\squishlist
    \item \textbf{Transparent to ad publishers:} As a third-party who pays the hosting website for displaying its ad contents, an ad publisher is generally reluctant to change the way they operate their services.
    \item \textbf{Easily deployable at the hosting website:} From the perspective of the hosting website, the process of injecting perturbations into the target page should be mostly automatic and convenient. We envision an additional procedure in website deployment via which the web pages will go through
    to make changes to the page.
\squishend

\subsection{Overview}
\point{Optimization problem formulation}
%
%
Formally, consider the optimization problem:
\begin{equation} \label{eq:opt}
\begin{aligned}
& \underset{x_{adv}}{\text{minimize}}
& & \mathrm{Dist}(x_{adv}-x_{input}) \\
& \text{subject to}
& & P_{model}(x_{adv}) \neq P_{model}(x_{input}), \\
&&& x_{adv} \in \mathcal{H}_{feature-space}, \\
&&& x_{adv} \in \mathcal{H}_{application-space},
\end{aligned}
\end{equation}
\noindent where, $Dist()$ measures the cost of adding the generated perturbation, $\mathcal{H}_{feature-space}$ and $\mathcal{H}_{application-space}$ denote the hyperspace that actionable examples can exist in
the feature space and application space, respectively.
%
Note that as discussed in \S \ref{sec-2},
it is hard for conventional $L_p$ norms to capture the real cost of adding a perturbation.
For this, we also modify the $L_{inf}$ norm to take the domain uniqueness into account, as discussed in the next subsection.

\begin{figure*}[htbp]
\centering
\begin{minipage}{0.53\textwidth}
  \includegraphics[width=\textwidth]{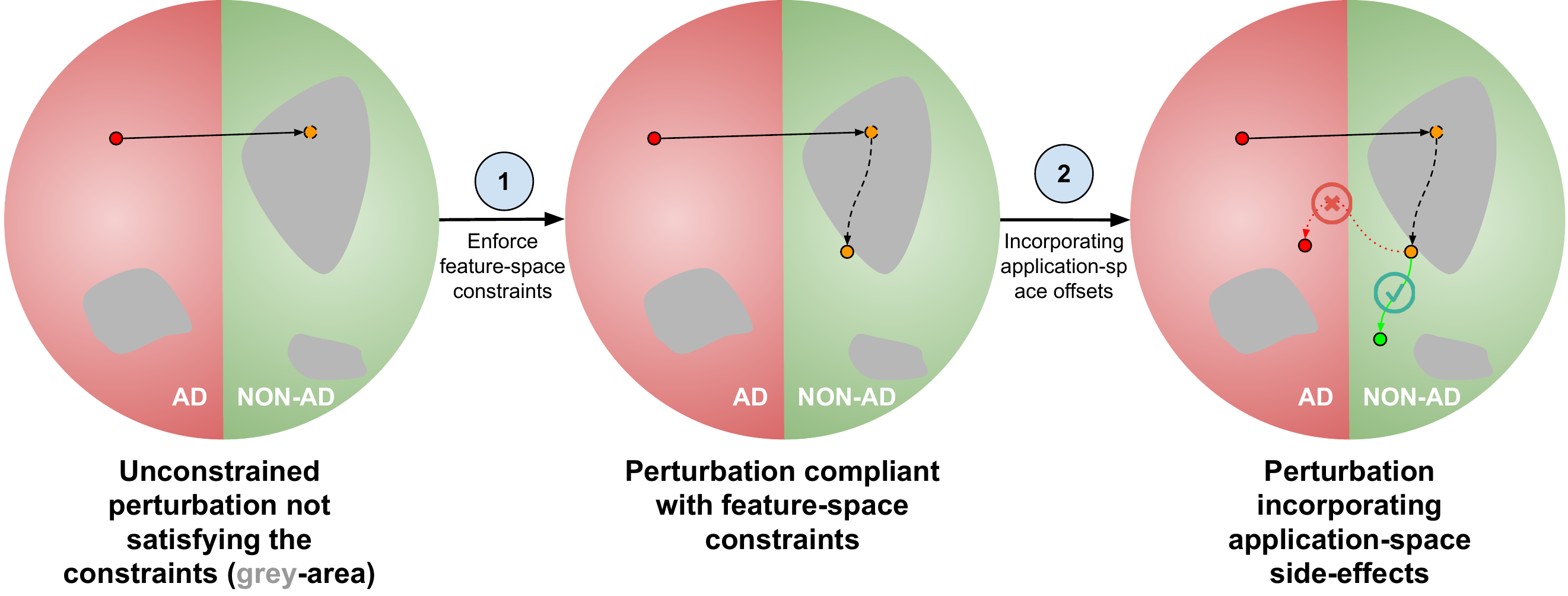}
  \caption{Perturbation trajectory in hyperspace for a search iteration}
  \vspace{-1em}
  \label{figure:hyperspace-trajectory}
\end{minipage}
\hspace{0.28in}
\begin{minipage}{0.28\textwidth}
  \includegraphics[width=\textwidth]{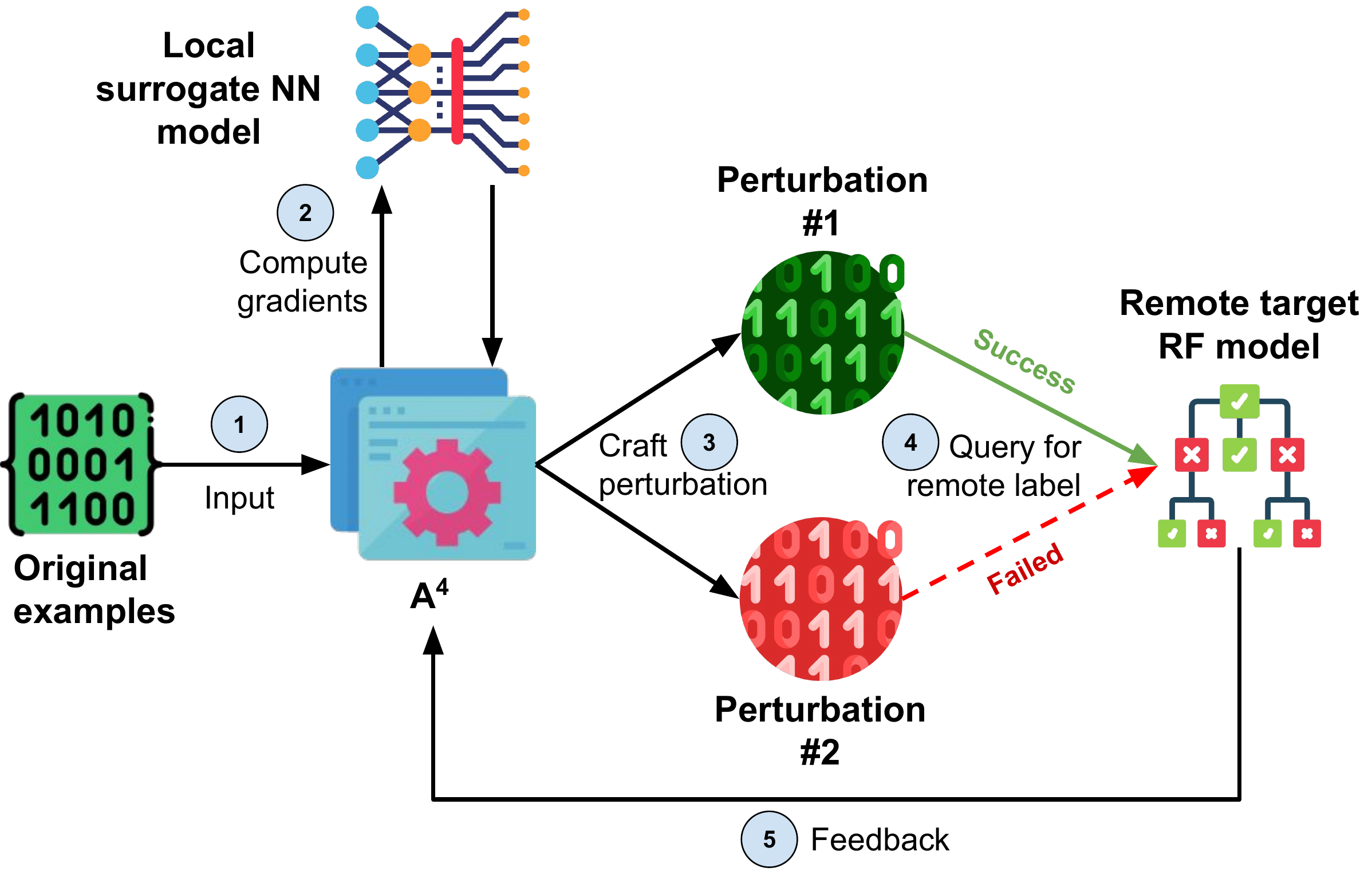}
  \caption{Transfer-based attack}
  \vspace{-1em}
  \label{figure:transfer-attack}
\end{minipage}
\end{figure*}

\point{Iterative search}
%
Since the optimization problem defined in Equation \ref{eq:opt} does not have an analytic solution \cite{bruna2013intriguing},
we instead approximate one iteratively through a search procedure.
%
captured in the pseudo-code in Algorithm \ref{algo:search-algo}.
%
\begin{algorithm}
\SetAlgoLined
\SetKwInOut{Input}{Input}\SetKwInOut{Output}{Output}
\SetKwFunction{FGen}{GenerateFeatureSpacePerturbation}
\SetKwFunction{FEnf}{EnforceFeatureSpaceConstraints}
\SetKwFunction{FMap}{MapBackToWebPage}
\SetKwFunction{FExt}{ExtractFeatureValues}
\SetKwFunction{FVer}{VerifyIfAdversarialOnTargetModel}
\Input{target model $M$, ad request $x_{input}$, maximum iterations $max\_iter$, maximum perturbation magnitude $\epsilon$}
\Output{actionable adversarial example $x_{adv}$}
$success \longleftarrow False$

$curr\_iter \longleftarrow 0$

$x_{curr} \longleftarrow x_{input}$

\While{$curr\_iter < max\_iter$ {\bf and} $success \neq True$} {
  $curr\_iter \longleftarrow curr\_iter + 1$

  $pert_{curr\_iter} \longleftarrow$ \FGen{$M, x_{input}, curr\_iter, \epsilon$}

  $x_{curr} \longleftarrow x_{input} + pert_{curr\_iter}$

  $x_{curr} \longleftarrow$ \FEnf{$x_{curr}$}

  $page_{perturbed} \longleftarrow$ \FMap{$x_{curr} - x_{input}$}

  $x_{curr} \longleftarrow$ \FExt{$page_{perturbed}$}

  $success \longleftarrow$ \FVer{$x_{curr}$}
}
\Return $x_{curr}$
\caption{\tool: \textbf{A}ctionable \textbf{A}d \textbf{A}dversarial \textbf{A}ttack}
\label{algo:search-algo}
\end{algorithm}
%
%
%
The search process not only enforces the feature-space constraints,
but also incorporates corrections to address application-space side-effects that occur when mapping feature-space perturbations back to the web application domain.
The key guiding principle is to
take small steps (in \textit{each iteration}) and corrective actions so that we are always on the right path. (details in the next subsection).
%
To better illustrate the framework, we show for each iteration, how the generated original perturbation is moved in hyperspace to ensure its actionability in Figure \ref{figure:hyperspace-trajectory}.	
The intuition is that errors may accumulate across multiple iterations and
can mislead us if we do not correct them at every step (and we take smaller steps
for the same reason).
Inspired by the iterative philosophy that underpins PGD,
\tool also divides the overall optimization problem into multiple iterations.

\point{Transfer-based attack}
%
%
To craft a successful grey-box attack, we need to use the dataset for training AdGraph to train a local \textit{surrogate} model that is differentiable, and then use this model to estimate the gradients and craft adversarial examples accordingly.
These type of attacks are considered ``transfer-based'' because the successful adversarial examples crafted locally need to be adversarial on a remote target model that is different and possibly unknown.
We depict \tool's transfer-based attack generation in Figure \ref{figure:transfer-attack}.
Prior research \cite{papernot2016transferability} has shown that this so-called inter-model transferability exist in almost all modern ML models (including non-differentiable ones such as Random Forest).
\point{Perturbable feature selection}
Before delving into constraints, we need to first manually identify what features to perturb.
These features must be perturbable, indicating that the attacker must know how to map
the perturbations from the feature space back to concrete changes in the application space, i.e., the web page.
As mentioned in \S \ref{sec-2}, AdGraph has two categories of features: structural and content-based (URL-related).
Generally, we follow three principles to pick features: (i) \textit{high impact}; (ii) \textit{ease of manipulation} and (iii) \textit{compliance of contraints} (as defined in the next subsection).
Towards satisfying (ii) and (iii), we start from URL features and pick 6 of them (\#2-\#7 in Table \ref{table:perturbed-features}) as they intuitively indicate the "ad-ness" and are relatively easy to tweak (simple string manipulations for keyword addition/removal). 
%
%
To further guarantee (i), we run ablational experiments by perturbing differnt combinations of features, and show that features \#2-\#7 provide quite limited evasion rates. 
Specifically, these URL features only offer a success rate of 37.10\% (which as shown later 
is $\approx 1.6 \times$ lower than what is achieved with \tool with all its features).
Thus, as a key novelty in \tool, we add structural feature \#1 that is designed to encode the size of the graph. In order to avoid affecting ad functionality of the request (part of principle (iii)), we limit our pertubation on \#1 to being additive only.
Under the principles, after systematically analyzing all the 65 features used in AdGraph, we identify 7 seed features from both (URL and structural) categories that the hosting website of the ad request can conveniently control and perturb in practice.
Table \ref{table:perturbed-features} shows their semantics and data types.
\subsection{Feature-Space Constraint Enforcement}
In this subsection, we describe how \tool imbibes explicit numeric constraints
in the feature space.
%
%
%
Since the constraints defined in the feature space can be generally considered for three main purposes - validity, functionality and stealthiness, 
we need to enforce them differently.

\point{Validity constraints}
These constraints keep the perturbed features numerically valid i.e.,
they guarantee that they fall within meaningful \textit{domains of definitions}.
For instance, features \#1 and \#7 are counts of nodes and characters,
and cannot be negative or non-integers;
binary features \#2 to \#6
should always take values of a 0 or a 1.
These constraints are enforced by projecting any perturbed value falling outside the meaningful domain of definition back to the domain.
Concretely, we define three projection operations for binary and integer types of perturbable features in $\texttt{EnforceFeatureSpaceConstraints()}$ in Algorithm \ref{algo:search-algo}:
%
\begin{equation} \label{equation:proj}
x_{proj}=
\begin{cases}
\text{min}(0, x_{pert}), & \text{if } x \in S_{integer}\\
0, & \text{if abs} ((x_{pert} - 0) \leq (x_{pert} - 1)) \\
   & \text{and } x \in S_{binary}\\
1, & \text{if abs} ((x_{pert} - 0) > (x_{pert} - 1)) \\
   & \text{and } x \in S_{binary}
\end{cases}
\end{equation}
Through these operations, the perturbed features of our choice are guaranteed to be valid in the feature space.

\point{Functionality constraints}
Besides validity, \tool also needs to ensure that the generated feature-space perturbations won't
break any functionality of the original ad request.
Hence, we also enforce functionality constraints onto the perturbed features.
To do so, we follow two principles which we
refer to as \textit{non-decreasing} and \textit{semantic equivalence}.
For counter-like features like \#1 and \#7, we limit their perturbed values to be greater than or equal to the original values;
otherwise, we project the modified value back to its original value.
We refer to this as the ``non-decreasing principle.''
This projection reflects our assumption that adding information to the web page should not break any existing functionality, but removing existing items might harm the semantics in an unpredictable fashion.

\begin{table}[b]
\centering
\fontsize{9}{9}\selectfont
\begin{tabular}{c|c|c}
\hline
\textbf{No. \#} & \textbf{Meaning} & \textbf{Data Type} \\ \hline
1 & \begin{tabular}[c]{@{}c@{}}Total number of nodes in the graph\\ at the time of classification\end{tabular} & Integer \\ \hline
2 & \begin{tabular}[c]{@{}c@{}}If predefined ad keywords appear \\ in the URL string\end{tabular} & Binary \\ \hline
3 & \begin{tabular}[c]{@{}c@{}}If predefined special characters \\ appear in the URL string\end{tabular} & Binary \\ \hline
4 & \begin{tabular}[c]{@{}c@{}}If any semicolon appears in \\ the URL string\end{tabular} & Binary \\ \hline
5 & \begin{tabular}[c]{@{}c@{}}If the base domain of the current\\  page appear in the URL string\end{tabular} & Binary \\ \hline
6 & \begin{tabular}[c]{@{}c@{}}If predefined ad dimension keywords\\  appear in the URL string\end{tabular} & Binary \\ \hline
7 & Length of request URL & Integer \\ \hline
\end{tabular}
\caption{Perturbed features}
\label{table:perturbed-features}
\vspace{-2em}
\end{table}

For URL features \#2 to \#7, we need to ensure that after perturbations, the original functionalities/semantics of the request are still preserved.
We leverage the fact that for HTTP(s) requests, different character encoding schemes end up delivering the same information to the ad server,
as long as the ad server can decode the messages properly.
%
In this case, for features \#2 to \#6 that detect predefined keywords/characters from the URL string,
we can simply change the default ASCII encoding to HTML encoding.
Since AdGraph assumes all URL text to be ASCII encoded, our perturbations can bypass its detection over all URL-related features.
For feature \#7, we choose to append random characters to increase its value, and place the appended string as an unused query,
which avoids disrupting other functional parts of the URL.
Note that these URL manipulations introduce changes to the request received by the server,
ane therefore might require cooperation from the server-side (e.g. support of different text encodings).
We argue such cooperation is rather convenient and fits our threat model mentioned previously ---
the hosting website and third-party ad publishers are motivated to slightly change their server
configurations to bypass adblockers and recover revenues.
We refer to this latter process as preserving ``semantic-equivalence.''

For clarity, we also define the enforcement operations for functionality constraints formally as follows:
\begin{equation} \label{eqution:functionality-const}
x_{proj}=
\begin{cases}
\text{max}(x_{input}, x_{pert}), & \text{if } x \in S_{count-based}\\
\text{re-encode}(x_{pert}), & \text{if } x \in S_{URL-based}
\end{cases}
\end{equation}

\point{Stealthiness constraints}
Besides validity and functionality, the generated perturbations should also achieve a high level of stealthiness.
Specifically, the perturbations that \tool applies on features will have to be limited by a threshold.
Conventionally, the perturbation size is measured via the use of $L_p$ norms.
However, these norms are unsuitable for AdGraph's feature set. First, with many binary/categorical features, use of $L_p$ norms blindly treats all features as having the same scale, which is not the case in reality.
For example, changing a binary feature from 0 to 1 means that the status it represents has flipped. This is fundamentally different from an integer feature changing by the same amount; for the latter, it could
indicate
that its real value has changed from a minimum to a maximum value (due to data normalization happened in dataset pre-processing).
%
Thus, if we set a threshold $L_{inf}$ to 0.3, binary features can never be flipped (as the flipping threshold is 0.5), whereas integer values can still change even if a normalization is applied.
To avoid such scale mis-interpretations, we propose a customized $L_{inf}$ norm which is defined as follows.
\begin{equation} \label{eqution:custo-norm}
L_{cust\_inf}(x)=
\begin{cases}
\text{max}(\text{abs}(x_i)), & \text{if } x_i \in S_{numeric}\\
x_i, & \text{if } x_i \not\in S_{numeric}
\end{cases}
\end{equation}

Besides customizing the norm, we also slightly modify the operation
for clipping a perturbation within the norm.
Specifically, conventional clipping functions (e.g. the one used by PGD) regard the global range of a particular feature across the whole dataset as the base of the clipping threshold, for conventional features.
For web pages, such clippings can easily lead to overly large perturbations as the ranges of many numeric features can vary very drastically from website to website.
Therefore, we change the clipping from relying on a global range to a local per-website range, as formally defined in Equation \ref{eqution:clipping}
\begin{equation} \label{eqution:clipping}
Clip'(x_i; \epsilon_{g}, \epsilon_{l}) = \begin{cases}Clip(x_i; \epsilon_{g}*r_i), & \text{if }\epsilon_{g}*r_i < \epsilon_{l}*x_i\\ Clip(x_i;\epsilon_{l}*x_i), & otherwise\end{cases} 
\end{equation}
where $\epsilon_{g}$ is the global threshold, $\epsilon_{l}$ is the local threshold, $r_i$ is the global range of $x_i$ with respect to this particular feature in the training dataset, given by $x_{imax} - x_{imin}$, and $Clip(x_i, \epsilon)$ is the standard clipping operation defined in Equation \ref{equation:clipping}.
As shown in \S \ref{sec-5}, our customized norm along with the localized clipping operation, helps limit the effective size of generated perturbations, and thus improves the stealthiness significantly compared to the traditional setup of $L_{inf}$ norms and global clipping.

\subsection{Application-Space Side-Effect Incorporation}
Now that we have generated feature-space adversarial perturbations that comply with manually-defined domain constraints,
we need to map them back to concrete changes in the web page representations.
As discussed previously, ideally these perturbed feature values should all be reflected accurately
in the page.
This can be overt if we can re-extract the feature vector from the perturbed web page and
verify that it
matches the expected one.
%
%

However, introducing changes (e.g. total number of nodes) to the web page can bring
about unpredictable offsets to values of other features
that are not included in the feature-space perturbations.
Specifically, there are several inter-dependent features considered by AdGraph such as \texttt{average\_degree\_connectivity}.
As we add perturbations nodes to the page to perturb the feature counting the total number of nodes in the graph, feature \texttt{average\_degree\_connectivity} might also \textit{nondeterministically} change as the maximum per-node connection degree is raised, 
which might end up turning an adversarial perturbation into non-adversarial.
More critically, such feature value offsets/drifts are impossible to be prediceted, and therefore cannot be pre-computed in closed-loop formulas, which motivates our design of executing the feedback loop.
%

\point{Feedback loop}
%
%
\begin{figure*}
\centering
\begin{minipage}{0.55\textwidth}
  \includegraphics[width=\textwidth]{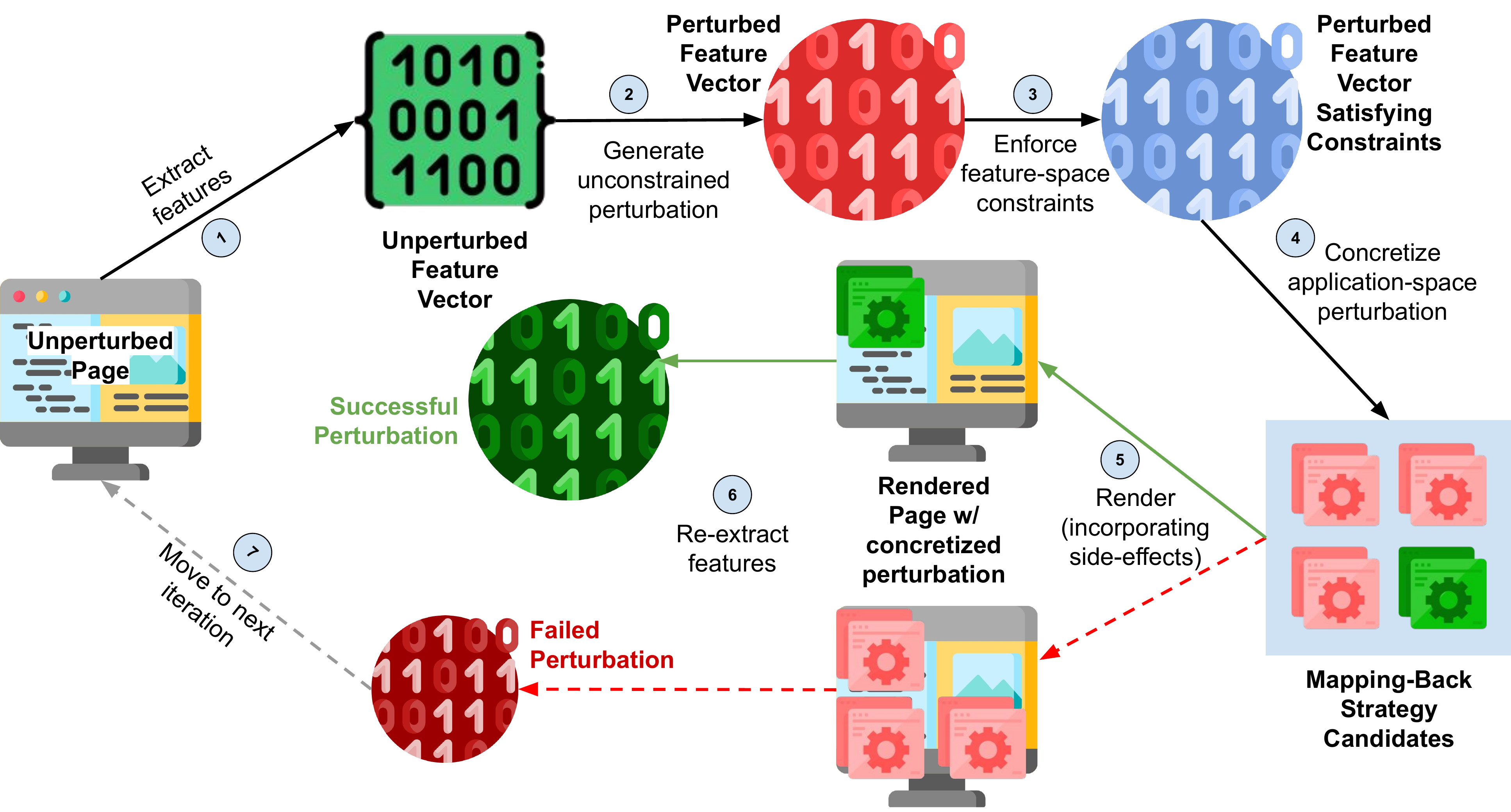}
  \caption{Proposed feedback loop in \tool's each search iteration}
  \label{figure:feedback-loop}
\end{minipage}
\begin{minipage}{0.3\textwidth}
  \centering
  \includegraphics[width=\textwidth]{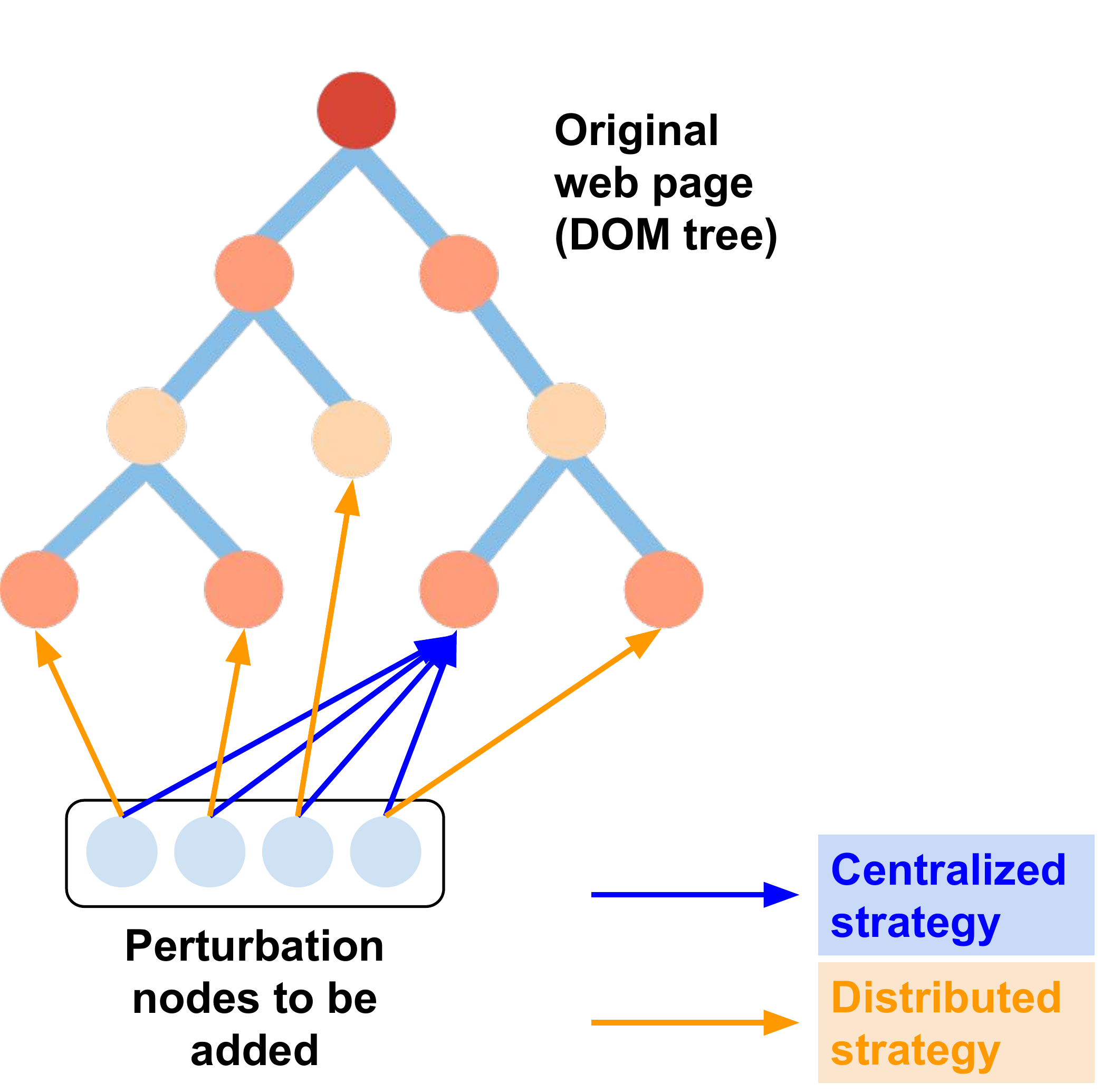}
  \caption{Different mapping-back strategies on adding perturbation nodes}
  \label{figure:mapping-back}
\end{minipage}
\end{figure*}
To incorporate such unpredictable side-effects,
we \emph{passively observe} how changes in one feature leads to changes in others.
Specifically, we first map the controllable feature-space perturbations back to the web pages
by ``rendering'' the page in a lightweight fashion (we use the timeline structure in Chromium~\cite{adgraphCode}) to do so 
and then re-extract all the features to \textit{capture} the side-effects.
We
finally verify if the final perturbation (with side-effects) can still evade detection.
If so, we are done; else, we continue the iterative search procedure to find another candidate perturbation
(we enlarge the current step size by a step size to generate a new gradient).
Effectively, we have created an automated feedback loop as illustrated in Figure \ref{figure:feedback-loop}.

\point{Mapping-back strategies}
For some features, there are multiple ways to concretize the feature-space perturbations as changes to web pages
(step 4 in Figure \ref{figure:feedback-loop}).
For instance, there are multiple ways to increase the total number of nodes in a page (feature \#1 in Table \ref{table:perturbed-features}).
We can choose to place these nodes either as the children of a single existing node (\textit{centralized} strategy), or as the children of multiple existing nodes (\textit{distributed} strategy), as shown in Figure \ref{figure:mapping-back}.
These different mapping-back strategies introduce different side-effects to the feature values,
and can hence affect the effectiveness of the final adversarial example (as depicted by the red and green points in Figure \ref{figure:hyperspace-trajectory}).
One example is that for the feature \texttt{average\_degree\_connectivity}, the centralized strategy is likely to lower the feature value significantly after
the map-back as the added nodes cause crowding and thus, raise the current maximum number of connections per node in the graph; this is the denominator in the formula that computes \texttt{average\_degree\_connectivity}.
In contrast, the distributed strategy tends to have negligible side-effects with respect to this feature.
In order to maximize the chance of finding a successful adversarial example, we apply all feasible mapping-back strategies in the feedback loop, and then verify their results.
This helps \tool discover as many green point cases as possible (Figure \ref{figure:hyperspace-trajectory}).

%% file: docs/evaluation.tex
\section{Evaluation}
\label{sec-5}
In this section, we first evaluate \tool's effectiveness in terms of its success rate and algosithm convergence in crafting adversarial examples against AdGraph; then, we analyze the generated perturbations from several perspectives; finally, we assess \tool's performance overhead \footnote{We will open source the implementation of \tool's attack pipeline with its datasets for reproducibility and for future extensionsi, at the time of publication.}.

\point{Dataset}
Since \tool primarily targets the current version of AdGraph, we need to first replicate the classification performance reported in \cite{iqbal2020adgraph} to ensure that our evaluation is sound.
In order to do so, we reached out to the authors of \cite{iqbal2020adgraph} and reproduced the classification pipeline that they used for AdGraph.
Given that the web crawl conducted and presented in \cite{iqbal2020adgraph} was in early 2018, and is therefore outdated, 
we carried out a new crawl on September 15th, 2019 to collect the graph representation of the landing pages of Alexa's top 10k websites.
Then, we processed these graphs and extracted 65 features to form the dataset ready for ML tasks.
Table \ref{table:dataset-stats} lists some basic statistics from the crawled dataset.
\begin{table}[]
\centering
\fontsize{9}{9}\selectfont
\begin{tabular}{l|c}
\hline
\textbf{Stat} & \textbf{Value} \\ \hline
\begin{tabular}[c]{@{}l@{}}\# successfully crawled\\ records (distinct requests)\end{tabular} & 586,218 \\ \hline
\begin{tabular}[c]{@{}l@{}}\# successfully crawled \\ websites (distinct domains)\end{tabular} & 8,121 \\ \hline
\# features before one-hot encoding & 65 \\ \hline
\# features after one-hot encoding & 332 \\ \hline
\# categorical features & 4 \\ \hline
\# binary features & 36 \\ \hline
\# numeric features & 25 \\ \hline
\begin{tabular}[c]{@{}l@{}}\# records to perturb \\ (distinct requests)\end{tabular} & 1,430 \\ \hline
\end{tabular}
\caption{Dataset statistics}
\label{table:dataset-stats}
\vspace{-2em}
\end{table}

From the 586,218 request records,
we pick 60,000 as the test set (i.e. the remaining 526,218 records are used as a training set). 
These are used to test the accuracy of trained classifiers and craft adversarial examples.
To showcase \tool's effectiveness over high-impact hosting websites, 
we select resource requests collected from Alexa's top 500 sites to be included in testing set,
from which we randomly pick 1,430 unique requests as the target ad resources to be perturbed.

\point{Model training}
To replicate the target classifier in \cite{iqbal2020adgraph}, we use the popular open-source machine learning library \texttt{scikit-learn}
to train a Random Forest (RF) model based on the crawled training dataset.
This is then used as the target model that \tool queries, for each given perturbed example, to verify the attack result.
We show the model's hyper-parameters and classification accuracy metrics over the partitioned testing set in Table \ref{table:rf-details}.
\begin{table}[]
\centering
\fontsize{9}{9}\selectfont
\begin{tabular}{l|c}
\hline
\textbf{\# trees} & 100 \\ \hline
\textbf{Split criterion} & entropy \\ \hline
\textbf{\begin{tabular}[c]{@{}l@{}}Maximum\\ tree depth\end{tabular}} & unlimited \\ \hline
\textbf{Precision} & 0.79 \\ \hline
\textbf{Recall} & 0.82 \\ \hline
\textbf{Accuracy} & 0.94 \\ \hline
\end{tabular}
\caption{Remote RF's hyper-parameters and accuracy metrics}
\label{table:rf-details}
\vspace{-1em}
\end{table}
As depicted, the accuracy metrics with our reproduced RF model are close enough to the one reported in Table \ref{table:rf-details}, which validates
our replication effort.

We need a local surrogate model that is differentiable (recall \S \ref{sec-4}), 
to drive our gradient-based attack.
To this end, we use a 3-layer Neural Network (NN) to be the surrogate model; we point out that a NN is considered to have the best model capacity \cite{papernot2016transferability} for imitating the decision boundaries of other models.
The hyper-parameters and accuracy metrics of this NN are in Table \ref{table:nn-details}. Note that in order to best mimic the remote decision boundary, we use the dataset that trains the target RF classifier and labels given by the \textit{target model}, instead of ground-truth labels, to train the local NN. Hence, the accuracy in Table \ref{table:nn-details} represents the agreement rate between two models.
\begin{table}[]
\centering
\fontsize{9}{9}\selectfont
\begin{tabular}{l|c}
\hline
\textbf{\# hidden layers} & 3 \\ \hline
\textbf{\# neurons} & (1024, 512, 128) \\ \hline
\textbf{Dropout rate} & 0.1 \\ \hline
\textbf{Accuracy (agreement rate)} & 0.91 \\ \hline
\end{tabular}
\caption{Local NN's hyper-parameters and agreement rate} 
\label{table:nn-details}
\vspace{-1em}
\end{table}

\point{Baseline attacks}
For comparison, we consider two baseline attacks we describe below (the descriptions
also illustrate the necessity of our proposed solution).
\squishlist
    \item \textbf{Weak baseline:} In this attack, we apply the standard PGD and do not enforce any feature-space constraints, other than the basic perturbation size limit ($\epsilon$ in Equation \ref{equation:pgd} and \ref{equation:clipping}) and a basic domain of definition (i.e., $0.0<x_i<1.0$).
    \item \textbf{Strong baseline:} In this attack, we apply \tool for one iteration only (i.e., enforce constraints and execute the feedback loop once), instead of performing iterative repetitions. 
    The purpose of this setup is to validate the benefits/advantages of the proposed iterative search framework.
    Without multi-iteration corrections, we anticipate difficulty in achieving high success rates, mainly because single-step approach might lead to application-space offsets that disrupt the adversarial nature of the example.
\squishend
Table \ref{table:param-attack} shows the hyper-parameters used with our weak/strong baselines and 
 \tool. 
Note that the parameter enforcement interval here refers to the number of steps we take in the gradient-based search before we enforce the constraints and call it one iteration of the feedback loop.
%
We operate at units of intervals instead of steps, because multiple steps are often required for binary features 
to be transitioned across the flipping threshold of 0.5, as explained in \S \ref{sec-4}.
%
%
Note that to avoid confusion, we use the term ``iteration'' to refer to an enforcement interval.
These parameters are empirically chosen, and we have tuned them to pick the best parameter set that are shown to yield best performances.
We also would like to point out that out of the different combinations of parameters, the improvement achieved by \tool over baseline attacks are generally consistent.
\begin{table}[]
\centering
\fontsize{9}{9}\selectfont
\begin{tabular}{l|c}
\hline
\textbf{Hyper-parameter} & \textbf{Value} \\ \hline
\begin{tabular}[c]{@{}l@{}}Iterations \\ ($max\_iteration$ in Algorithm \ref{algo:search-algo})\end{tabular} & 20 \\ \hline
Step-size & 0.07 \\ \hline
\begin{tabular}[c]{@{}l@{}}Maximum \textit{global} perturbation \\ threshold ($\epsilon_{g}$ in Equation \ref{equation:clipping}) \end{tabular} & 0.3 \\ \hline
\begin{tabular}[c]{@{}l@{}}Maximum \textit{local} perturbation \\ threshold ($\epsilon_{l}$ in Equation \ref{equation:clipping}) \end{tabular} & 0.8 \\ \hline
\begin{tabular}[c]{@{}l@{}}Enforcement interval\end{tabular} & 15 \\ \hline
\end{tabular}
\caption{Hyper-parameters used for attacks}
\label{table:param-attack}
\vspace{-1em}
\end{table}

\point{Success rate}
We summarize the results achieved by three attacks in terms of their success rates 
of finding actionable adversarial examples out of the 1,430 ad requests sampled from Alexa's top 500 websites, in Table \ref{table:attack-res}.
We see that \tool achieves the highest success rate in generating mis-classified examples while guaranteeing their actionability. It is almost twice as successful (84.3\% improvement) as the strong baseline.
This large margin of improvement shows the power of the iterative search adopted by \tool.
In comparison, the weak baseline fails to produce any valid perturbation because if none of the constraints is enforced, features of data types like binary or categorical can be changed into meaningless values (e.g. in a one-hot-encoded vector, more than one feature becomes 1). This makes it impossible for these
perturbed examples to be rendered in the browser at all.
%
\begin{table}[]
\centering
\fontsize{9}{9}\selectfont
\begin{tabular}{l|c|c}
\hline
\textbf{Attack} & \textbf{Success} & \textbf{Fail} \\ \hline
Weak baseline & 0 (0\%) & 1430 (100\%) \\ \hline
Strong baseline & 465 (32.52\%) & 965 (67.48\%) \\ \hline
\tool & \textbf{857} (\textbf{59.93\%}) & 573 (40.07\%) \\ \hline
\end{tabular}
\vspace{-1em}
\caption{Breakdown of attack results}
\label{table:attack-res}
\vspace{-1em}
\end{table}

\begin{figure}[htbp]
	\centering
	\includegraphics[width=0.8\columnwidth]{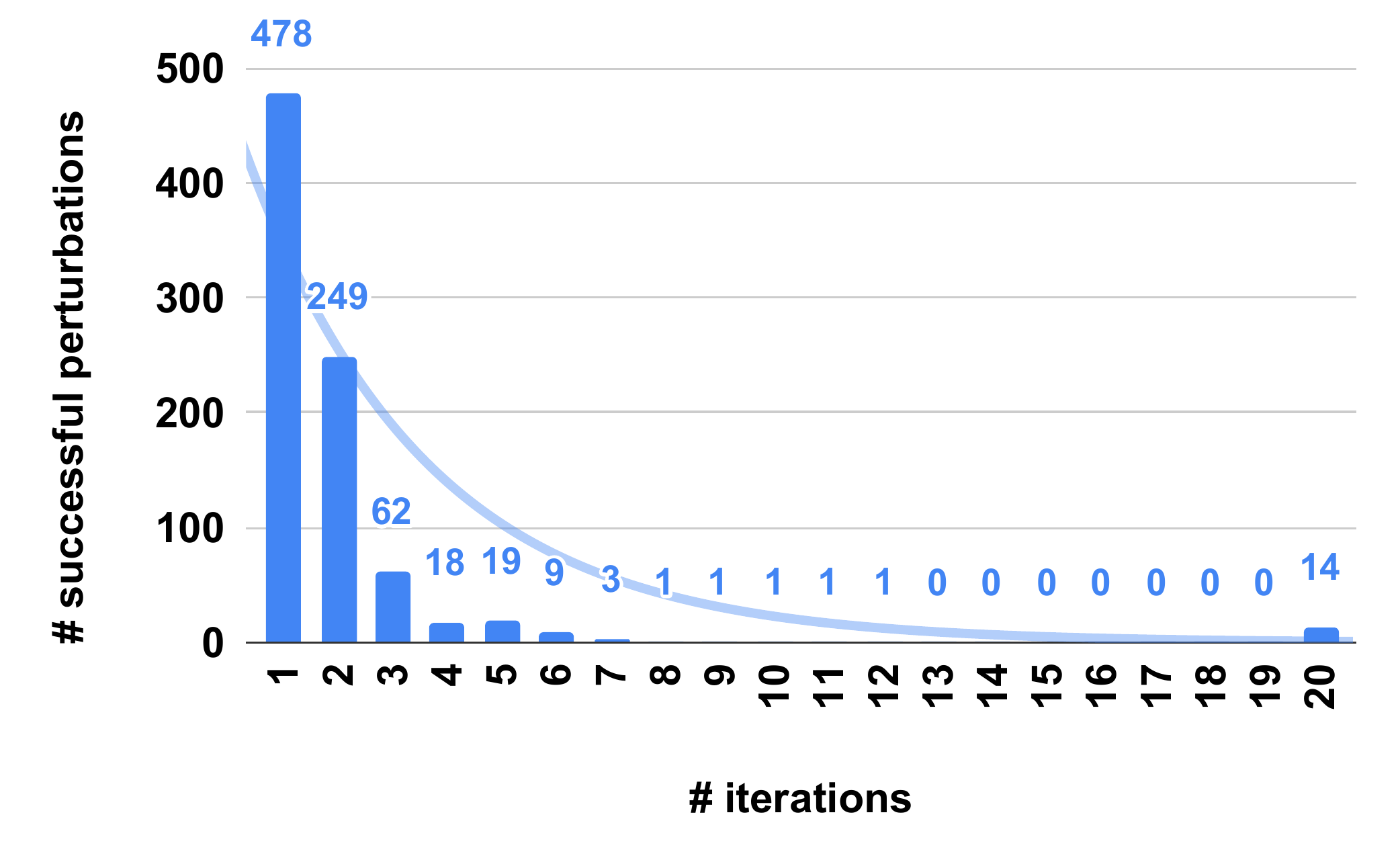}
	\caption{Attack convergence analysis}
	\vspace{-1em}
	\label{figure:attack-convergence}
\end{figure}

\point{Attack convergence}
In Figure \ref{figure:attack-convergence}, we show a histogram of the number of iterations a
that are needed to 
reach convergence for all the successful adversarial perturbations generated by \tool, 
We see that most (\textgreater90\%) cases converge within 5 iterations; this shows that the 
iterative enforcement of needed constraints and the feedback loop in \tool are 
extremely efficient in generating the adversarial examples.

\begin{figure*}
    \begin{minipage}{0.35\textwidth}
        \centering
        \includegraphics[width=\textwidth]{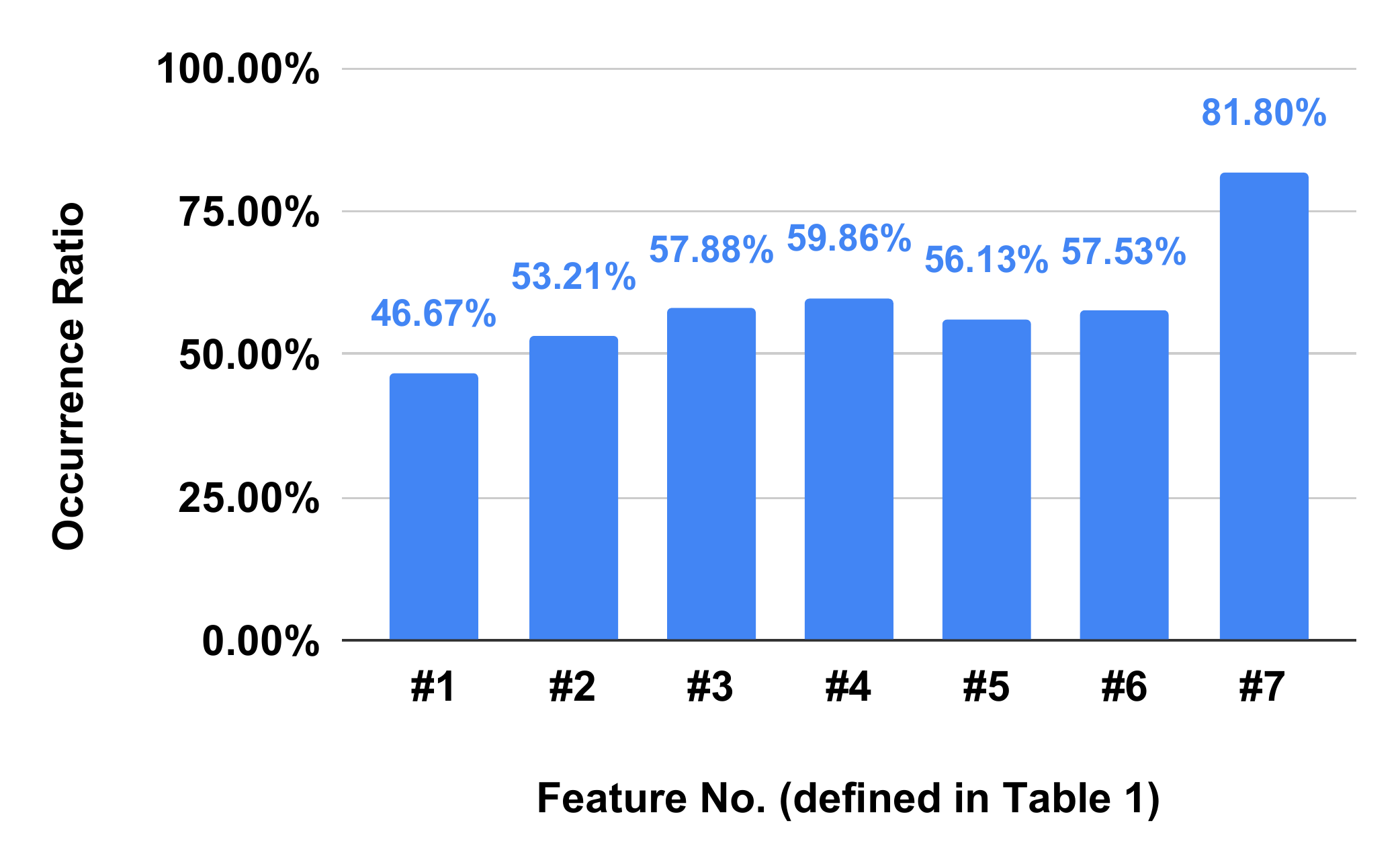}
        \caption{Feature significance analysis}
        \vspace{-2em}
        \label{figure:feature-sig}
    \end{minipage}
    \hfill
    \begin{minipage}{0.55\textwidth}
    \fontsize{9}{9}\selectfont
        \begin{tabular}{c|c|c|c|c|c|c|c}
            \hline
            \textbf{\begin{tabular}[c]{@{}c@{}}Feature\\ No.\end{tabular}} & \textbf{\#1} & \textbf{\#2} & \textbf{\#3} & \textbf{\#4} & \textbf{\#5} & \textbf{\#6} & \textbf{\#7} \\ \hline
            \textbf{Ave.} & \begin{tabular}[c]{@{}c@{}}865.06\\ (0.9\%)\end{tabular} & -0.08 & 0.17 & 0.48 & 0.40 & 0.44 & \begin{tabular}[c]{@{}c@{}}21164.71\\ (19.3\%)\end{tabular} \\ \hline
            \textbf{Max.} & \begin{tabular}[c]{@{}c@{}}13825\\ (14.5\%)\end{tabular} & n/a & n/a & n/a & n/a & n/a & \begin{tabular}[c]{@{}c@{}}32937\\ (30.0\%)\end{tabular} \\ \hline
            \textbf{Min.} & 0 & n/a & n/a & n/a & n/a & n/a & 0 \\ \hline
        \end{tabular}
        \captionof{table}{Per-feature perturbation statistics}
        \label{table:pert-stats}
    \end{minipage}
\end{figure*}

\point{Perturbation analysis}
To better understand the generated perturbations, we investigate (1) \textit{feature significance}, which shows for all successful examples, what features are modified more/less frequently in the generated perturbations;
(2) \textit{feature perturbation stats}, which capture the statistics of successful perturbations with respect to each feature; and
(3) \textit{mapping-back strategy significance}, which captures ``which mapping-back strategy is more likely to lead to a successful perturbation.''
We show the resulting histogram/tables of these analysis in Figure \ref{figure:feature-sig}, Table \ref{table:pert-stats} and \ref{table:mapback-sig}, respectively.

\begin{table}[]
\centering
\begin{tabular}{l|c|c|c|c}
\hline
 & \textbf{Total} & \textbf{Centr.} & \textbf{Distr.} & \textbf{Both} \\ \hline
\textbf{\#} & 857 & 66 (7.70\%) & 64 (7.47\%) & 727 (84.83\%) \\ \hline
\end{tabular}
\caption{Mapping-back strategy significance analysis}
\label{table:mapback-sig}
\vspace{-2em}
\end{table}
%


From Figure \ref{figure:feature-sig}, we observe a high level of significance (above or close to 50\%) associated with all features. 
This indicates their significant importance in crafting successful perturbations.
With the most significant feature being \#7, the length of the request URL, it's shown that increasing the length of ad requests helps raise their adversarial potential the most 
from among all the perturbed features.
Note that this seemingly contradicts the conclusion reached in \cite{iqbal2020adgraph}, which suggests that longer URLs predict higher ``ad-ness'' from the overall data distribution.
We argue that these two observations are actually compatible: for adversarial examples, we are essentially exploiting the discrepancies between target model's decision boundary and the real patterns that can accurately distinguish ad and non-ad requests, bounded by the constraints and perturbation cost budget.
The overall data distribution presented in \cite{iqbal2020adgraph} does not define the local discrepancy, or perturbation space with respect to a particular example.
The existence of such adversarial areas in hyperspace exactly showcases the vulnerability that ML models rely too much on overall statistical cues to make classification decisions.


Table \ref{table:pert-stats} summarizes the statistics of the perturbed features.
Note that the percentages appended below the statistics of numeric features \#1 and \#7 are their relative ratios with respect to the data ranges from the crawled dataset, indicating their relative changes (either increase or decrease) compared to the original values.
Through these ratios, we can see that by customizing the $\L_{inf}$ norm and enforcing per-page limits 
(recall \S \ref{sec-4}), \tool has shrunken the size of the generated perturbations for numeric features (i.e. \#1 and \#7). 
%
For example, for feature \#1, on average, only 0.9\% additional nodes (with respect to the maximum total number of nodes in the dataset) are added to the page to flip the classification result.
Smaller number of perturbation nodes (added) brings two advantages: (1) better stealthiness of the attack, 
since (a) it is harder for adblockers to differentiate the perturbation from normal nodes, when there are fewer dummy nodes and (b) these nodes are properly obfuscated against simple detection as \tool does in its implementation; and (2) lower performance overhead for page loads.

Table \ref{table:mapback-sig} reports the significance of different mapping-back strategies that \tool tried in its feedback loop.
As can be seen, in more than 80\% of the successful perturbations, 
both centralized and distributed mapping-back strategies are attempted (to a similar degree) to craft actionable adversarial examples.
However, in close to 15\% of the test cases, only one strategy succeeds.
These cases show the advantage of applying multiple strategies to cope with the unpredictable application-space side-effects into the iterative search procedure used in \tool.
Essentially, more valid green points (as depicted in \ref{figure:hyperspace-trajectory}) can be discovered with additional strategies.


\point{Performace overhead}
Lastly, we report the performance overheads with \tool. This comprises of two components: (i) offline feature-space pertubration computation and (ii) online perturbation loading/rendering.
For (i), each example takes 5.8 min on average (from 1,430 ad requests).
We argue this is reasonable for an offline pre-computation setup in exchange for recovering ad revenue.
For (ii), rendering generated perturbations (adding 39 nodes) for a typical ad request \url{http://securepubads.g.doubleclick.net/gpt/pubads_impl_2019121002.js} on \url{https://www.kompas.com/} incurs an average overhead of 0.23 sec (from 100 loads), and is hence, minuscule compared to the original page load time 20.9s.

%% file: docs/discussion.tex
\section{Discussions and Limitations}
\label{sec-6}
\point{Generalizability}
Although \tool primarily targets an ML-based adblocker at this point, we would like to argue that its iterative search procedure and feedback loop are general enough to be applied to other AML scenarios in the web domain, or even other domains.
At a high level, any ML task that (1) fits the feature-/application-space paradigm as the generated examples should be actionable in both spaces, (2) is constrained in feature-space due to the validity/functionality requirements as discussed in \S \ref{sec-4}, and (3) has associated side-effect offsets upon mapping from feature-space to application-space (as shown in Figure \ref{figure:hyperspace-trajectory}), can be seen to be compatible with \tool's methodology.

For example, there are similar characteristics for ML-based malware classifiers.
First, most learning-based malware detectors still rely on feature extraction,
leaving the space of the discrepancy between these two spaces to be exploitable by \tool.
Second, common feature sets used by ML malware classifiers include a large portion of constrained feature types,
such as categorical (e.g., file type) and binary (e.g., whether a specific permission is requested~\cite{grosse2017adversarial,hu2018black,yuan2014droid}).
%
These constraints can be properly defined and enforced using the operations proposed by \tool.
Lastly, different features used by malware detectors can also be inter-dependent.
For example, changing the file type can lead to a different set of allowable permissions
(e.g., system app can request more permissions than regular apps).
This can also be incorporated leveraging the feedback loop in \tool.
To summarize, we anticipate it to be a promising direction to apply \tool's framework to other applications with adjustments, and leave it as future work.

\point{Improving ML-based adblocking}
Given the evaluation results and multi-dimensional analysis conducted in \S \ref{sec-5}, we would like to summarize some useful insights to improve the design and implementation of AdGraph, and other ML-based adblockers for detecting more ads and trackers accurately and robustly, amid rising attempts from ad publishers to cloak their ads.
First, ML-based adblockers should in general shed more light into their feature set selection.
During our analysis of AdGraph's features, for example, we find that there are several global features (e.g. \#1 in Table \ref{table:perturbed-features}) that encode the overall size of the constructed graph.
%
Since these features do not describe anything local with respect to the request node being classified, they are of lower importance, but leave unnecessary perturbation space for adversarial attacks.
Evidentally, the feature significance analysis presented in \S \ref{sec-5} and information gain from the Random Forest used in \cite{iqbal2020adgraph} both verify the relatively low predicative/distinguishing power of these global features.

%
Second, as URLs are the most functional part of a HTTP(s) request, adblockers should make more efforts in parsing them in order to prevent string manipulation tricks.
For AdGraph, our analysis shows that it does not handle HTML's special character encoding very well, which opens room for the stealthy perturbation of URL-related features.
More generally, adblockers are advised to implement more sanitizations to reduce their chances of taking adversarial perturbation as normal contents as much as possible, or put less weight on the URL-related features.

\point{Enhancing hosting websites for evasion}
For hosting websites that have incentives to fight against the emerging ML-based adblockers, our investigations suggest three aspects they can adjust with regards to their pages, to make it easier to mount adversarial
subversions against learning-based adblockers: (1) maintaining a sufficiently high level of page complexity to guarantee adequate space and flexibility for placing adversarial perturbations; if there are too few elements in the original web page, adversarial perturbations are naturally more detectable; (2) providing the necessary setups (e.g. proxy/rotating servers) to accommodate perturbations on URL-related features. This aligns with the common practice nowadays adopted by some websites countering rule-based adblockers \cite{zhu2018measuring}.

%
\point{Completeness of \tool's implementation}
As noted in \S \ref{sec-4}, currently \tool is designed to perturb only 7 features from the possible 64.
Although this conservative limit makes our attack stealthier in practice, we do plan to explore perturbations on more features in the future and further analyze their effectiveness.
Moreover, our current \tool implementation only has two mapping-back strategies (centralized and distributed)
as discussed in \S \ref{sec-4}.
While we argue that even with these two strategies, we can already showcase the
power of our proposed feedback loop,
we will explore additional strategies in the future to expand our search space.

%% file: docs/conclusions.tex
\section{Conclusions}
\label{sec-6}
In this paper, we present the design and implementation of \tool (\textbf{A}ctionable \textbf{A}dversarial \textbf{A}d \textbf{A}ttack), a new adversarial attack targeting the state-of-the-art learning-based adblocker AdGraph.
Unlike previous work on generating adversarial samples on unconstrained domains, \tool, explicitly
accounts for constraints that arise in the context of the web domain.
We show promising results in this unique domain which can have substantive implications in online advertising.
%
%
%